\documentclass[11pt,preprint]{aastex63}
\usepackage{graphicx} 
\usepackage{url}

\newcommand{\gtorder}{\mathrel{\raise.3ex\hbox{$>$}\mkern-14mu\lower0.6ex\hbox{$\sim$}}}
\newcommand{\ltorder}{\mathrel{\raise.3ex\hbox{$<$}\mkern-14mu\lower0.6ex\hbox{$\sim$}}}

\shorttitle{19.5 Hz QPO in GRB 211211A}
\shortauthors{Chirenti, Dichiara, Lien, Miller}

\begin{document}
\title{Evidence for a strong 19.5 Hz flux oscillation in Swift BAT and Fermi GBM gamma-ray data from GRB 211211A}

\correspondingauthor{Cecilia Chirenti}
\email{chirenti@umd.edu}

\author[0000-0003-2759-1368]{Cecilia Chirenti}
\affiliation{Department of Astronomy, University of Maryland, College Park, 20742, MD, USA.}
\affiliation{Astroparticle Physics Laboratory, NASA/GSFC, Greenbelt, 20771, MD, USA}
\affiliation{Center for Research and Exploration in Space Science and Technology, NASA/GSFC, Greenbelt, 20771, MD, USA}
\affiliation{Center for Mathematics, Computation and Cognition, UFABC, Santo Andre, 09210-170, SP, Brazil.}

\author[0000-0001-6849-1270]{Simone Dichiara}
\affiliation{Department of Astronomy and Astrophysics, The Pennsylvania State University, 525 Davey Lab, University Park, 16802, PA, USA.}

\author[0000-0002-7851-9756]{Amy Lien}
\affiliation{Department of Chemistry, Biochemistry, and Physics, University of Tampa, 401 W. Kennedy Blvd, Tampa, 33606, FL, USA.}

\author[0000-0002-2666-728X]{M.~Coleman~Miller}
\affiliation{Department of Astronomy and Joint Space-Science Institute, University of Maryland, College Park, MD 20742-2421 USA}

\begin{abstract}
The gamma-ray burst (GRB) GRB~211211A is believed to have occurred due to the merger of two neutron stars or a neutron star and a black hole, despite its duration of more than a minute.  Subsequent analysis has revealed numerous interesting properties including the possible presence of a $\sim 22$~Hz quasiperiodic oscillation (QPO) during precursor emission.  Here we perform timing analysis of Fermi and Swift gamma-ray data on GRB~211211A and, although we do not find a strong QPO during the precursor, we do find an extremely significant 19.5~Hz flux oscillation, which has higher fractional amplitude at higher energies, in a $\sim 0.2$~second segment beginning $\sim 1.6$~seconds after the start of the burst.  After presenting our analysis we discuss possible mechanisms for the oscillation.
\end{abstract}


\section{Introduction}

There are believed to be two basic categories of gamma-ray bursts \citep{1993ApJ...413L.101K}: those powered by a particular type of core-collapse supernova, which typically produce long bursts (durations of tens of seconds or larger), and those produced by the merger of two neutron stars or possibly a neutron star and a black hole, which typically produce short bursts (durations of a few seconds or shorter).  A growing number of gamma-ray bursts blur the lines between these categories. For example, the burst GRB~211211A lasted for more than a minute, yet the spectrum and especially the apparent presence of a kilonova after the burst suggest a merger rather than a core-collapse supernova \citep{2022Natur.612..223R,2022Natur.612..228T,2022Natur.612..232Y}.

Further elucidation of the nature of gamma-ray bursts in general, and anomalous bursts such as GRB~211211A in particular, could be obtained with the detection of quasi-periodic oscillations (QPOs) in the gamma-ray light curve.  For example, QPOs with frequencies $\nu\gtorder 1000$~Hz could be related to oscillations of a hypermassive neutron star or an accretion disk shortly after merger \citep{2019ApJ...884L..16C}; indeed, evidence for such oscillations was found in GRB~910711 and GRB~931101B by \citet{2023Natur.613..253C}.  

Lower-frequency oscillations, on the order of $\sim 10-100$~Hz, are also predicted in several models.  For example, if a neutron star merges with a rapidly spinning, low-mass black hole, then coherent Lense-Thirring precession of the resulting accretion disk could lead to QPOs in this frequency range \citep{2013PhRvD..87h4053S,2023arXiv230806151L}.

Here we report the detection of a strong oscillation, at a frequency $\nu\approx 19.5$~Hz, shortly after the beginning of the main part of GRB~211211A.  The oscillation is independently evident in both Swift and Fermi gamma-ray data.  Interestingly, a $\sim 22$~Hz QPO from the precursor of this same burst was reported by \citet{2022arXiv220502186X}; we see a power excess at this frequency during the precursor, but not with a high enough significance to claim detection.  The 19.5~Hz oscillation (which we will sometimes call a QPO although we do not formally resolve the frequency width of the oscillation) lasts for $\sim 0.2$~seconds, starts and ends abruptly, and has a higher fractional amplitude at higher photon energies.  Compared with a red-noise-only model, the Bayes factor in favor of a QPO is $\sim 6\times 10^{10}$ in the Swift BAT data alone, and $\sim 4\times 10^4$ in the Fermi GBM data alone.  We discuss different interpretations of this oscillation in Section~\ref{sec:discussion}, after presenting the description of our data in Section~\ref{sec:description} and our analysis in Section~\ref{sec:analysis}.

\section{Description of data and burst}
\label{sec:description}

GRB~211211A triggered the Swift Burst Alert Telescope (BAT; see \citealt{2005SSRv..120..143B}) at 13:09:59.634 UT on 2021 December 11 \citep{2021GCN.31202....1D}, and triggered the Fermi Gamma-ray Burst Monitor (GBM; see \citealt{2009ApJ...702..791M}) just 0.017 seconds later, at 13:09:59.651 UT on 2021 December 11 \citep{2021GCN.31201....1F}.  The burst was in the direction RA = 14h09m10.12s,
Dec = +27:53:18.1 (J2000), with an estimated redshift of $z=0.076$ ($D\approx 350$~Mpc) based on a galactic association \citep{2021GCN.31221....1M}.  The burst lasted more than a minute, but later association with a kilonova \citep{2022Natur.612..223R,2022Natur.612..228T,2022Natur.612..232Y} suggests that this was a long-duration ``short" GRB likely associated with the merger of two neutron stars or of a neutron star and a black hole.  Because we are interested in the timing properties of this burst, in this section we give details about how we extracted the Swift BAT and Fermi GBM data and how we time-align the data from the two satellites, as well as displaying the light curves and power spectra.

\subsection{Extraction of Swift BAT data}

For the QPO analysis, we created a 100 $\mu$s non-maskweighted light curve in $15-350$ keV using the BAT event data from the Swift BAT GRB catalog\footnote{\url{https://swift.gsfc.nasa.gov/results/batgrbcat/}}. The event data was created using the standard BAT GRB analysis tool, ``batgrbproduct\footnote{https://heasarc.gsfc.nasa.gov/ftools/caldb/help/batgrbproduct.html}'', version 2.48, which is part of the HEASoft analysis package. The 100 $\mu$s non-maskweighted light curve was created using the BAT analysis tool, ``batbinevt\footnote{https://heasarc.gsfc.nasa.gov/ftools/caldb/help/batbinevt.html}'', version 1.48. The non-maskweighted (i.e., non-background-subtracted) light curve was used because the QPO analysis in this work requires that the data obey Poisson statistics and do not have negative or fractional photon counts (which can be the case for a mask-weighted light curve).

\subsection{Extraction of Fermi GBM data}

To study the prompt emission of GRB 211211A we used the time-tagged event (TTE) data obtained from the two most illuminated sodium iodide (NaI) detectors (N2 and Na). We processed the data using the HEASoft \citep[version 6.30.1;][]{2014ascl.soft08004N} and the Fermitools software packages \citep[version 2.0.8;][]{2019ascl.soft05011F} following standard procedures \footnote{https://fermi.gsfc.nasa.gov/ssc/data/p7rep/analysis/scitools/gbm\_grb\_analysis.html}. Light curves were reconstructed using a 0.1~ms time bin, and considering different energy ranges ($8-1000$~keV, $4-37$~keV, $37-88$~keV, $88-166$~keV, and $>166$~keV) using the FSELECT and GTBIN tools.

\subsection{Time alignment of the Swift BAT and Fermi GBM data}
\label{sec:alignment}

Because the segment of interest is short and the 19.5~Hz signal appears to emerge and disappear suddenly, we align the starting times of the Swift BAT and Fermi GBM data prior to performing our analysis.

We start with the trigger times: as indicated above, the Fermi GBM trigger time is 0.017 seconds later than the Swift BAT trigger time.  We then note that at the time of trigger, the Fermi satellite was at longitude 197.17 degrees and latitude 24.57 degrees, at an altitude of 524 km \citep[derived using the Fermi GBM Data Tools;][]{GbmDataTools}, whereas the Swift satellite was at longitude 7.12 degrees and latitude of 20.36 degrees, at an altitude of 538 km.  Given the time, RA, and Dec of the burst, at the time of the burst it was above latitude $27.89^\circ$ and longitude $294^\circ$.  

To figure out the projection of the direction to the GRB onto the Fermi-Swift vector, we first compute the Fermi-Swift vector in Cartesian coordinates.  The average radius of the Earth is 6371 km, so at the time of the burst the distance from the center of the Earth to Swift was $r_S=6371+538=6909$~km and the distance from the center of the Earth to Fermi was $r_F=6371+524=6895$~km.  At the time of the GRB the colatitude for Swift was $\theta_S=90^\circ-20.36^\circ=69.64^\circ=1.2154$ radians and the azimuth for Swift was $\phi_S=7.12^\circ=0.1243$ radians.  Similarly, during the time of the GRB the colatitude for Fermi was $\theta_F=90^\circ-24.57^\circ=65.43^\circ=1.1420$ radians and the azimuth for Fermi wass $\phi_F=197.17^\circ=3.4413$ radians.  The three-dimensional locations of Swift and Fermi at the time of the burst were then
\begin{equation}
\begin{array}{rl}
{\rm Swift}&=r_S(\sin\theta_S\cos\phi_S,\sin\theta_S\sin\phi_S,\cos\theta_S)=(6427,803,2404)\,{\rm km}\\
{\rm Fermi}&=r_F(\sin\theta_F\cos\phi_F,\sin\theta_F\sin\phi_F,\cos\theta_F)=(-5991,-1851,2866)\,{\rm km}\; .\\
\end{array}
\end{equation}
Therefore the Fermi-Swift vector at the time of the burst was
\begin{equation}
{\rm Swift-Fermi}=(12418,2654,-462)~{\rm km}\; .
\end{equation}
The direction to the burst at the time of the burst had $\theta_{\rm GRB}=90^\circ-27.89^\circ=62.11^\circ=1.0840$ radians and $\phi_{\rm GRB}=294^\circ=5.1313$ radians.  The projected Fermi-Swift distance along the direction to the GRB equals the dot product of the unit vector toward the GRB, with the Fermi-Swift vector calculated above; positive means that the signal reached Swift first, whereas negative means that the signal reached Fermi first.  The unit vector is
\begin{equation}
{\hat\Omega}_{\rm GRB}=(\sin\theta_{\rm GRB}\cos\phi_{\rm GRB},\sin\theta_{\rm GRB}\sin\phi_{\rm GRB},\cos\theta_{\rm GRB})=(0.3595,-0.8074,0.4678)
\end{equation}
and the dot product of this with the Fermi-Swift vector is 2105 km, which is a light travel time of 0.007 seconds.  Subtracted from the 0.017 second difference in trigger times, this means that to align the Swift BAT and Fermi GBM light curves we shift the Fermi count arrival times by 0.01 seconds compared with their nominal values.

\subsection{Light curves, power spectra, and energy dependence}
\label{sec:lightpower}

\begin{figure}[h!]
    \centering
        \includegraphics[width=0.32\linewidth]{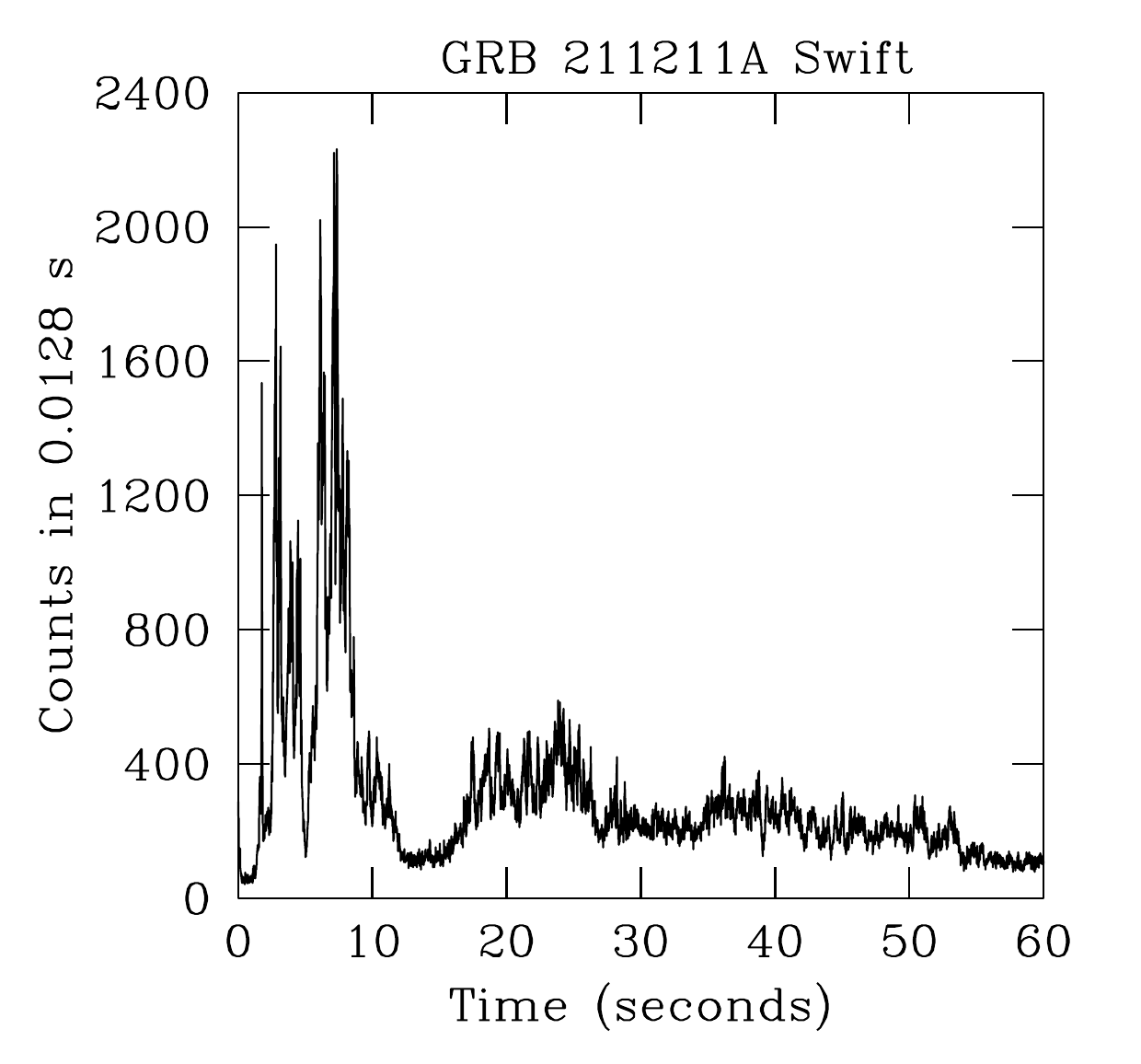}
        \includegraphics[width=0.32\linewidth]{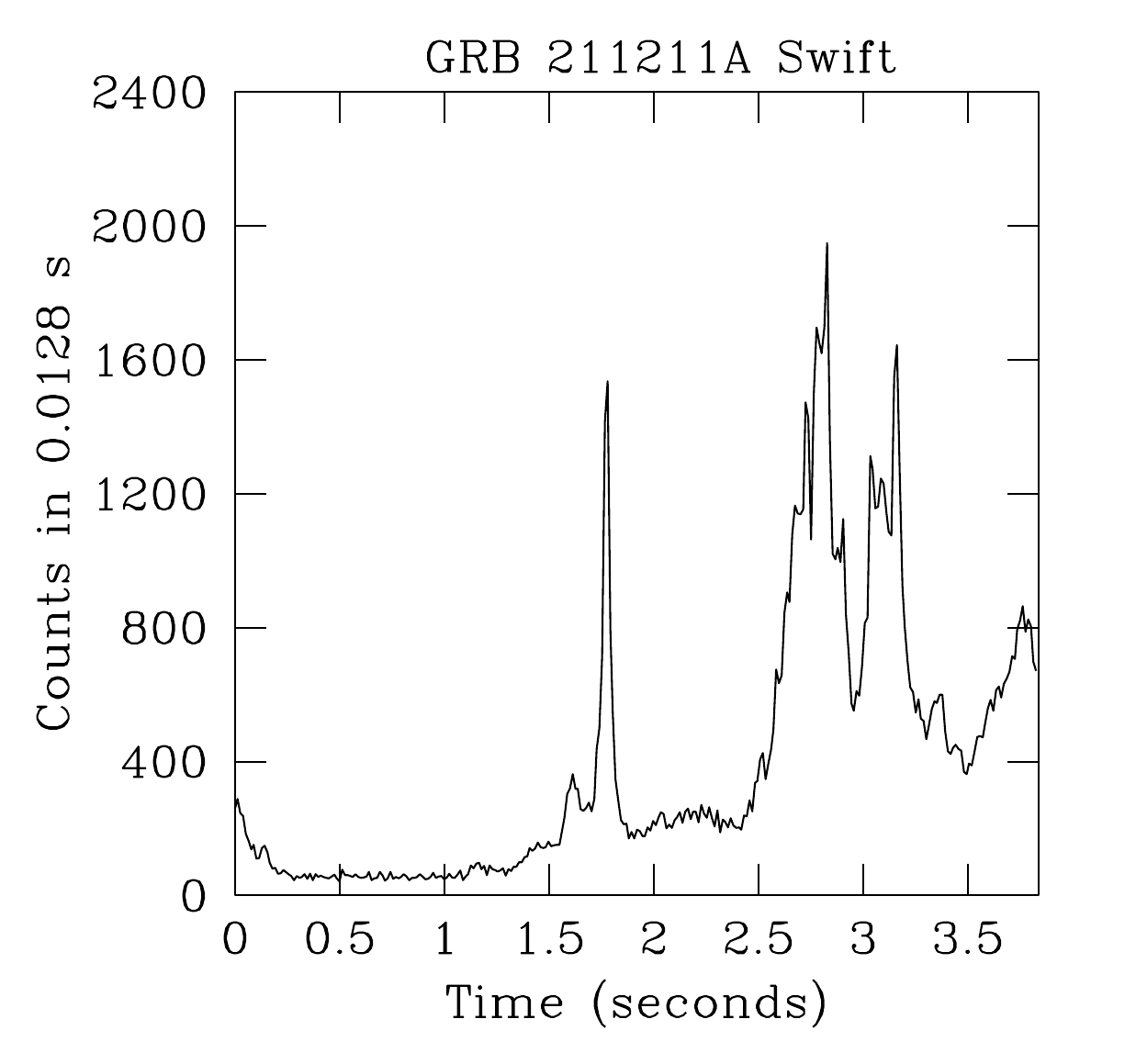}
        \includegraphics[width=0.32\linewidth]{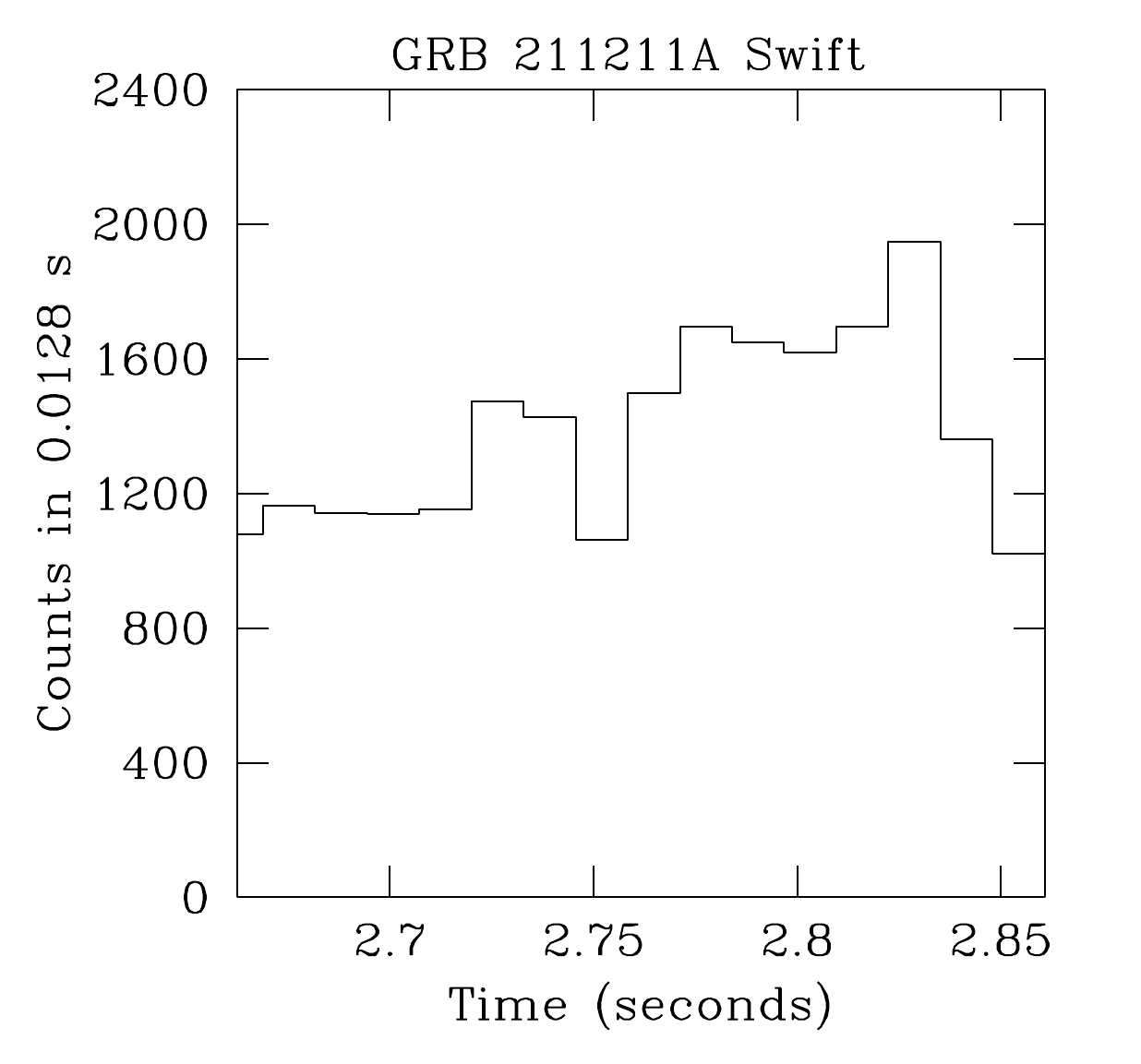}
        \includegraphics[width=0.32\linewidth]{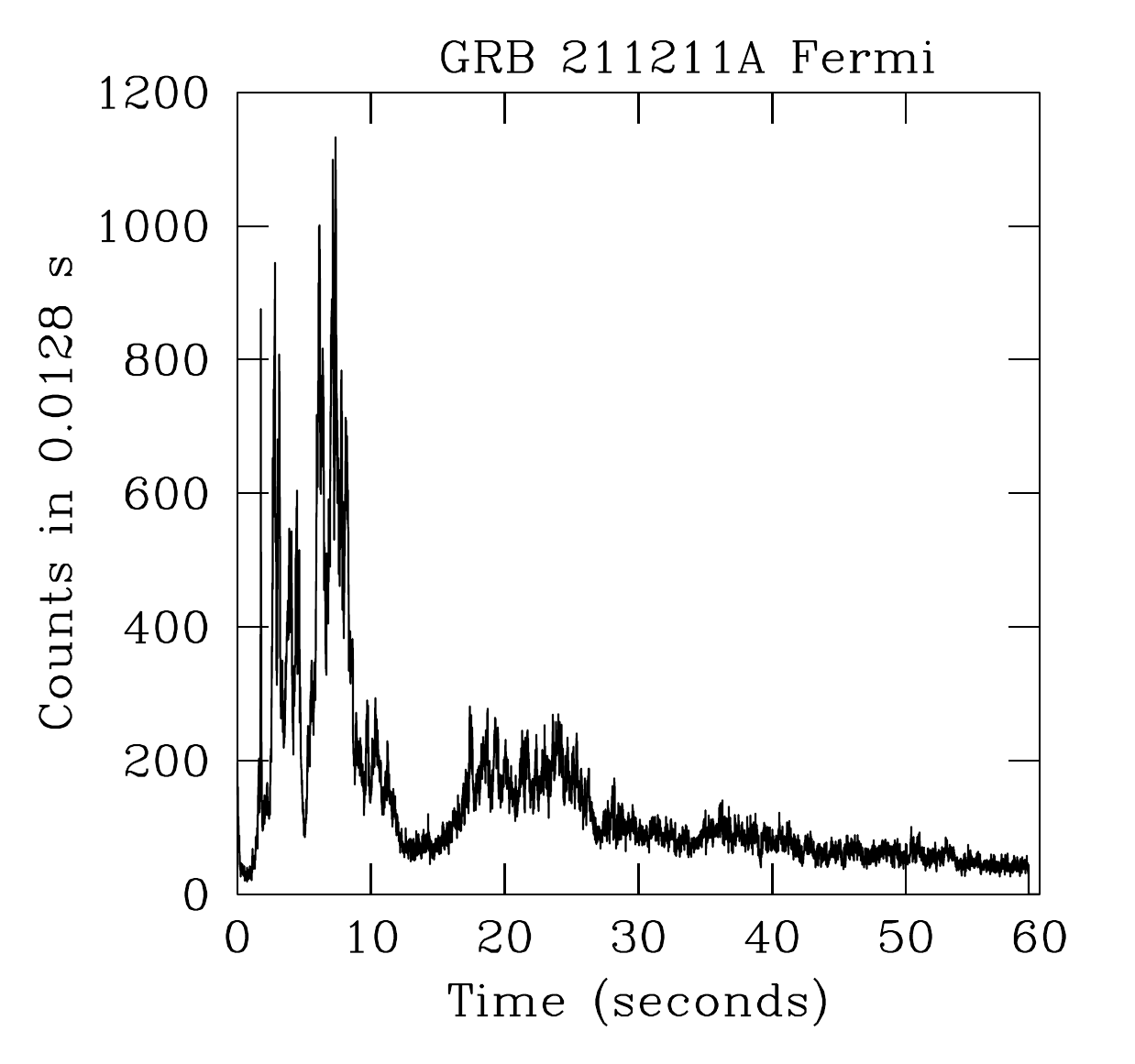}
        \includegraphics[width=0.32\linewidth]{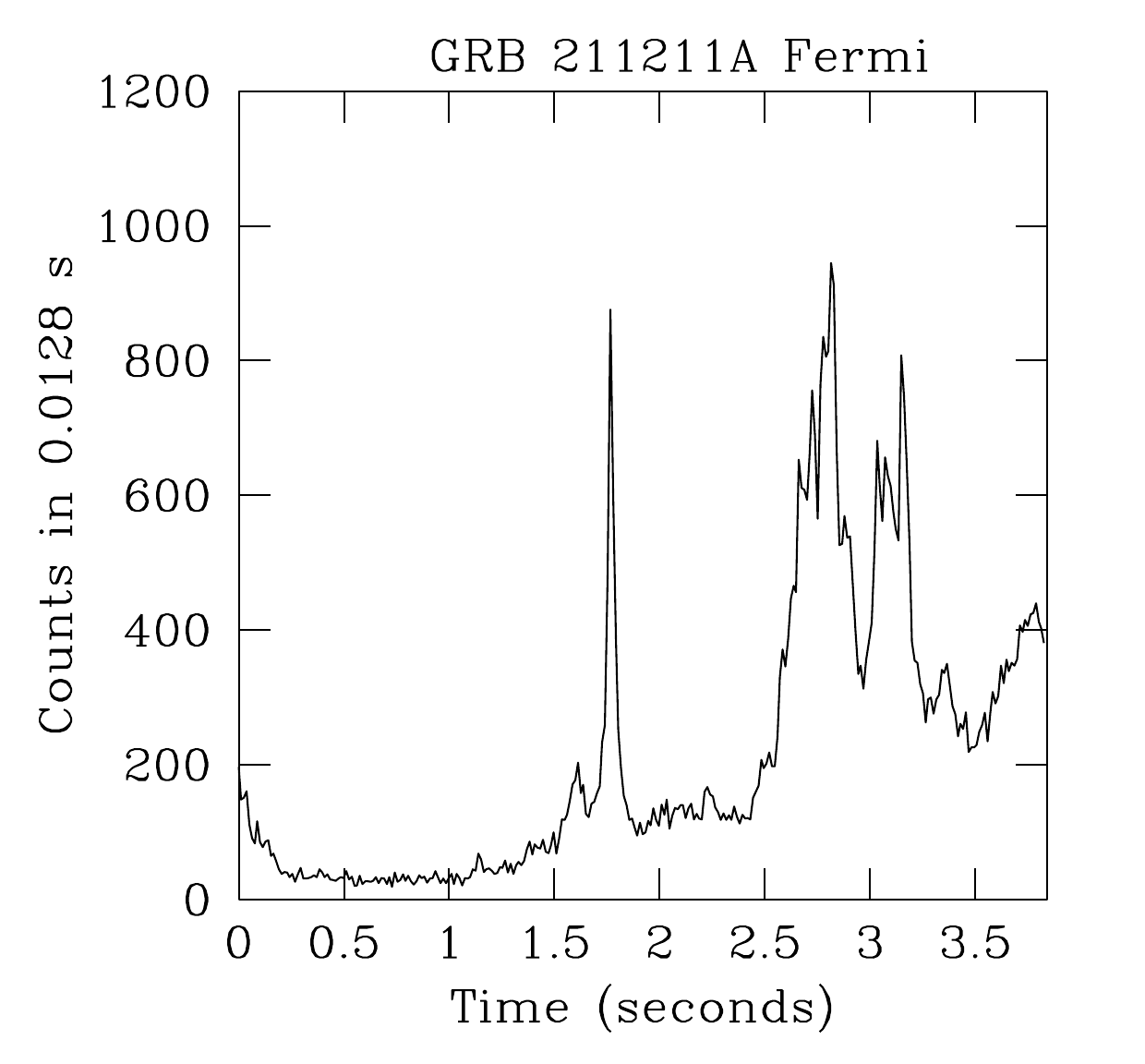}
        \includegraphics[width=0.32\linewidth]{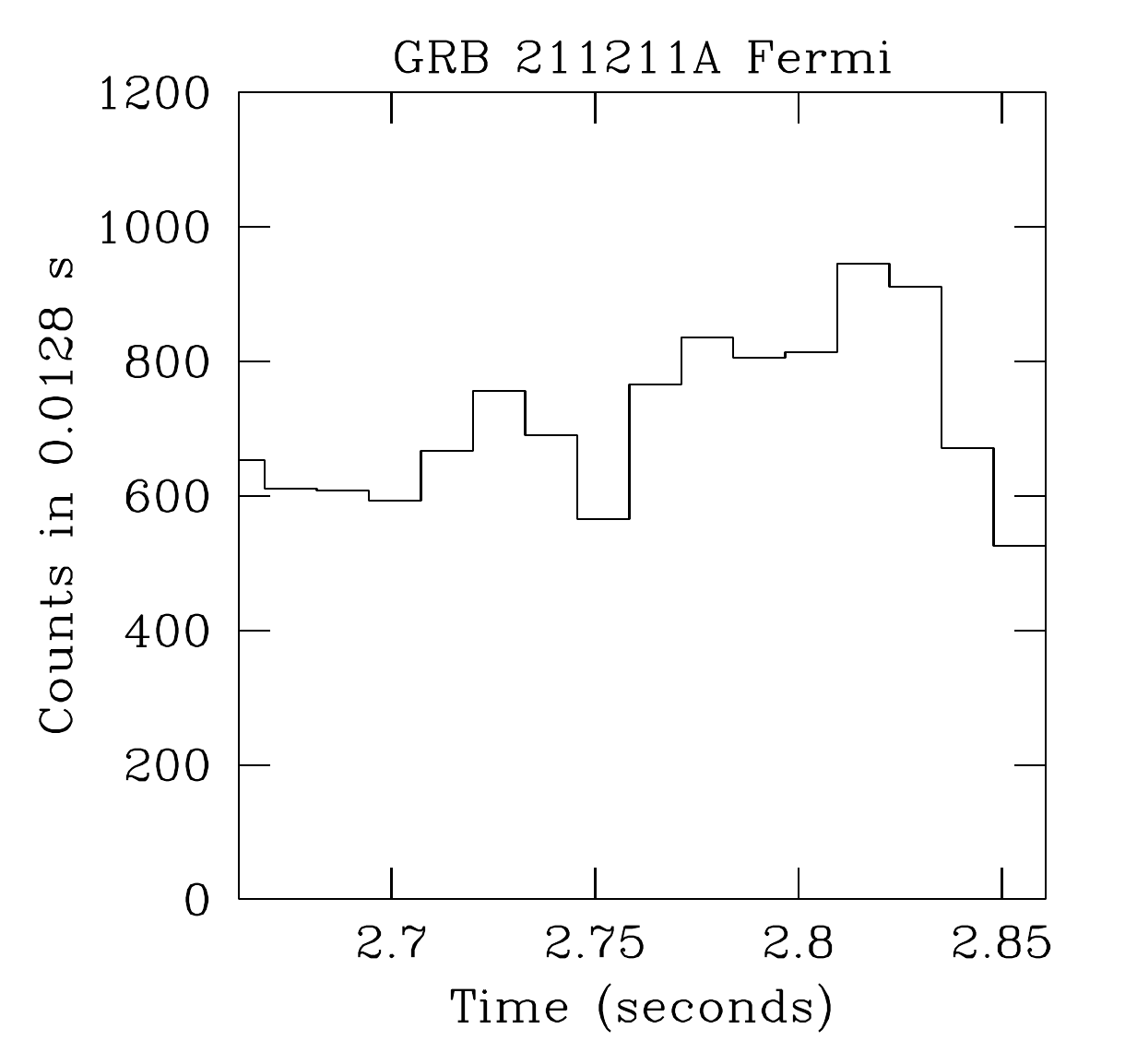}
    \caption{Light curves from Swift BAT (top panels) and Fermi GBM after the alignment procedure described in Section~\ref{sec:alignment} (bottom panels) over different time scales. In each case, we bin the data in intervals of 12.8~msec to make the development of the light curve more evident, although all of our analysis is performed using a time resolution of 0.1~msec.  The left, middle, and right panels show respectively the first minute of the burst, approximately the first four seconds (including the precursor at the beginning), and the 0.2048 second segment on which we focus our analysis.  The burst evidently has a long and complex light curve, and there is a strong correlation of substructure between the Fermi GBM and Swift BAT light curves, although BAT registers $\sim 2\times$ the number of counts as GBM.}
    \label{fig:lightcurves}
\end{figure}

The light curve of GRB~211211A is highly complex, as is evident from the Swift BAT and Fermi GBM light curves in Figure~\ref{fig:lightcurves}.  Here, to show the structure more clearly, we show data binned to 12.8~msec intervals, although we use 0.1~msec intervals in our analysis.  The Fermi GBM curve has been shifted following the procedure described in Section~\ref{sec:alignment}, and after this shift the correlations between the curves are evident.   The right-hand panels show the segment that displays the strong $\approx 19.5$~Hz signal in both the Swift BAT and the Fermi GBM data, and the middle panel also shows the precursor to the burst, from which \citet{2022arXiv220502186X} reported a $\approx 22.5$~Hz QPO.

\begin{figure}[h!]
    \centering
\includegraphics[width=\linewidth]{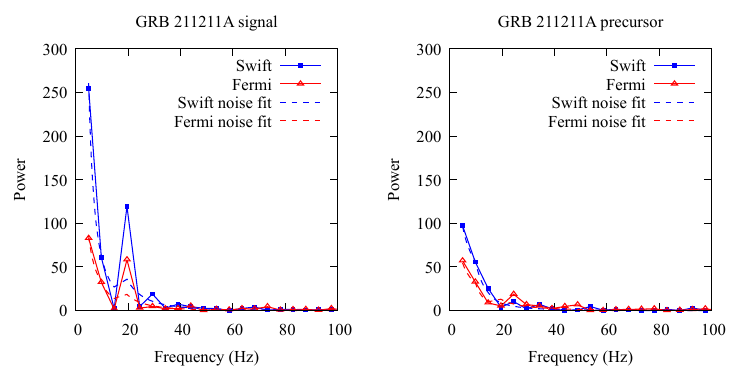}
    \caption{(Left panel) Power spectrum of our featured 0.2048-second segment.  Here we use the normalization from \citet{1975ApJS...29..285G}, in which the average power is 1 from an intrinsically constant signal with purely Poisson noise.  The frequency steps in the Fast Fourier Transform (FFT) used to produce this power spectrum are $1/0.2048~{\rm s}\approx 4.88$~Hz.  The dashed lines show the fits to the data sets using the multislope noise model described in Section~\ref{sec:analysis}, without a QPO; the noise-only model clearly underpredicts the power at 19.5~Hz.  The key feature that makes the signal stand out in our analysis is the high power at $\approx 19.5$~Hz, flanked by low powers at $\approx 14.6$~Hz and $\approx 20.4$~Hz.  This feature is seen independently in the Fermi GBM data (red lines and open red triangles) and in the Swift BAT data (blue lines and solid blue squares), which argues against an instrumental origin for this signal.  No other interval in this burst has such a strong feature.  (Right panel) Here, as a contrast to the left panel, we show the power spectrum from the 0.2048~second precursor to the burst.  The normalization and line/point types are the same as in the left panel.  This is the segment for which \citet{2022arXiv220502186X} reported a moderately significant $\sim 22.5$~Hz QPO.  There is indeed an excess of power near that frequency in both the Swift BAT and the Fermi GBM data, but it is much weaker than the signal we feature.}
    \label{fig:powerspectra}
\end{figure}

In Figure~\ref{fig:powerspectra} we see the power spectra, using Fermi GBM and Swift BAT data independently, for our 0.2048~second segment (left panel) and for the first 0.2048~seconds of the precursor (right panel), which is the segment from which \citet{2022arXiv220502186X} reported a $\approx 22.5$~Hz QPO.  In our featured segment the excess power is clear in both Swift BAT and Fermi GBM data at $\approx 19.5$~Hz, compared with the power at the next lower ($\approx 14.6$~Hz) and next higher ($\approx 20.4$~Hz) frequencies. The power spectra in Figure \ref{fig:powerspectra} show excess power in a single frequency (19.5 Hz), which therefore does not have a resolved width. The analysis in Section~\ref{sec:analysis} demonstrates that the excess power is significant for both sets of data even when compared with a flexible red noise model that can accommodate multiple slopes.  In contrast, although there is some excess power in the vicinity of the $\approx 22.5$~Hz signal noted by \citet{2022arXiv220502186X}, the significance is not high.

\begin{figure}[h!]
    \centering
        \includegraphics[width=0.49\linewidth]{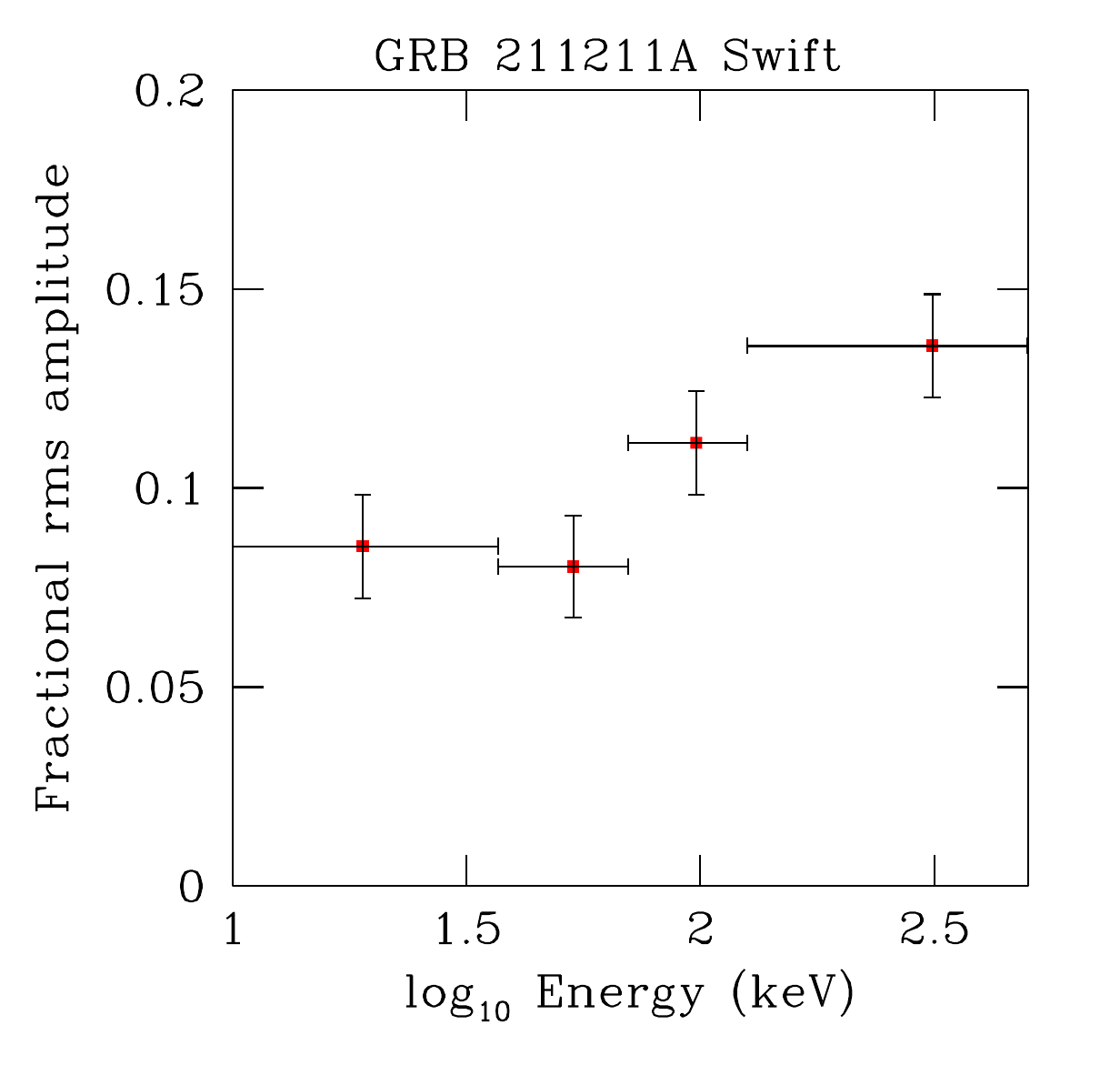}
        \includegraphics[width=0.49\linewidth]{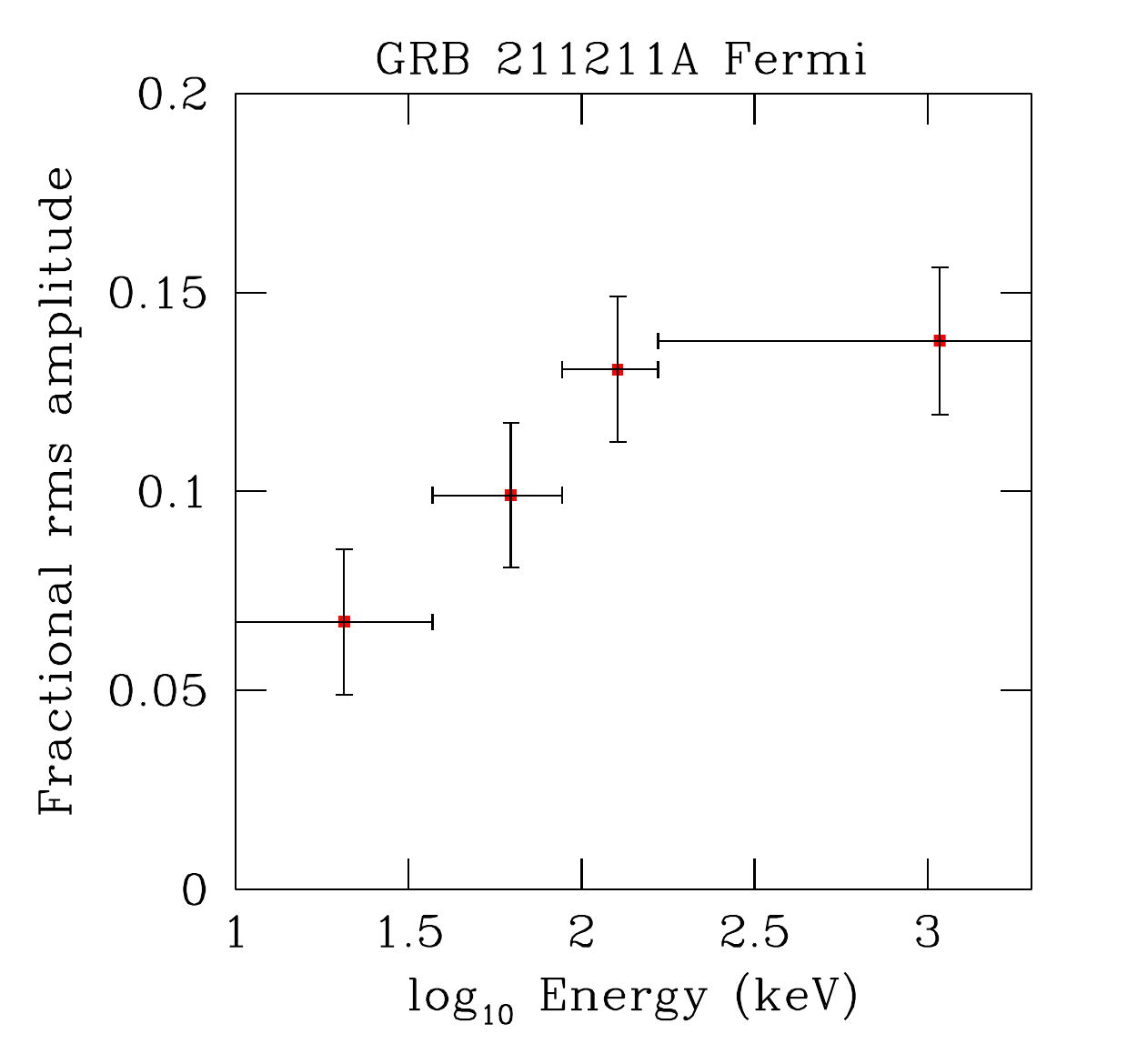}
    \caption{(left panel) Fractional rms amplitude of the $\approx 19.5$~Hz signal in four different energy ranges, selected to have approximately equal numbers of counts, from the Swift BAT data.  The energy ranges were roughly $<37$~keV, $37-70$~keV, $70-126$~keV, and $>126$~keV, up to a maximum energy of about 500~keV, and are indicated by the horizontal bars.  For each energy range the solid red squares indicate the median of the estimated amplitude and the vertical error bars indicate the $\pm 1\sigma$ ranges of the amplitude, as inferred using the power distributions discussed in \citet{1975ApJS...29..285G}; see text for details.  (right panel) The same, for the Fermi GBM data.  The energy ranges were roughly $4-37$~keV, $37-88$~keV, $88-166$~keV, and $>166$~keV, up to a maximum energy of about 2000~keV.  We see that in both data sets there is a clear increase in fractional rms amplitude with increasing energy.}
    \label{fig:ampenergy}
\end{figure}

In Figure~\ref{fig:ampenergy} we see the fractional rms amplitude versus energy for the Swift BAT data (left panel) and for the Fermi GBM data (right panel).  For each data set we break the data into four energy ranges with approximately equal numbers of counts.  For the Swift BAT data the energy ranges were roughly $<37$~keV, $37-70$~keV, $70-126$~keV, and $>126$~keV, up to a maximum energy of about 500~keV.  For the Fermi GBM data the energy ranges were roughly $4-37$~keV, $37-88$~keV, $88-166$~keV, and $>166$~keV, up to a maximum energy of about 2000~keV.  The vertical location of the solid red square in each energy range is the median amplitude (which, using our power normalization, equals $\sqrt{2P/N_{\rm counts}}$ for a power $P$ with $N_{\rm counts}$ counts) estimated using equation (16) of \citet{1975ApJS...29..285G}, and the upper and lower error bars show the $\pm 1\sigma$ amplitude using the same equation.  For both data sets, the amplitude rises with energy.

\begin{figure}[h!]
    \centering
        \includegraphics[width=0.49\linewidth]{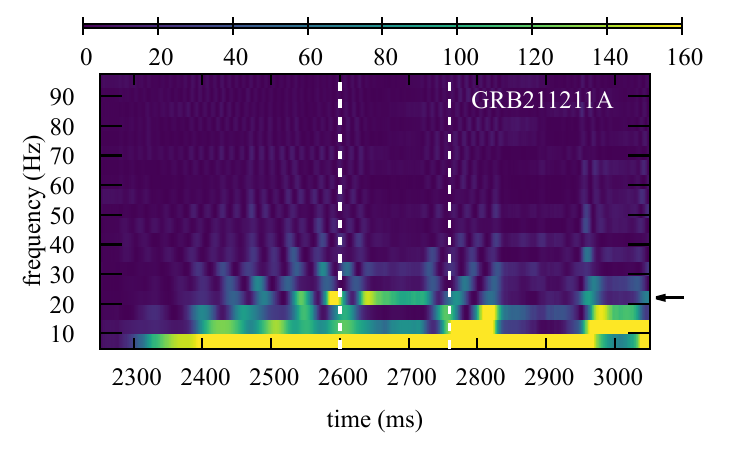}
        \includegraphics[width=0.49\linewidth]{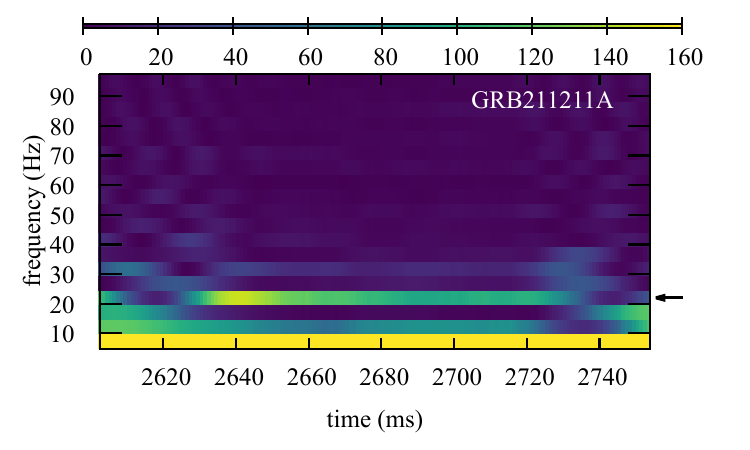}
    \caption{(left panel) Spectrogram of the full segment in the Swift BAT data of GRB 211211A that shows a strong 19.5 Hz signal. We use a 0.2048 sec window and shift it by 1 ms to cover the full segment. The power scale is saturated to match the highest power shown in the right panel. (right panel) Same as the left panel, but highlighting the strong signal.}
    \label{fig:spectrogram}
\end{figure}

The dynamical power spectra shown in Figure \ref{fig:spectrogram} allow us to estimate the approximate duration of the 19.5 Hz signal. We can see excess power in the signal frequency from approximately 2500 ms to 2700 ms after $T_0$. The strongest part of the signal, highlighted in the right panel, lasts for about 100 ms.    

\section{Analysis methods and results}
\label{sec:analysis}

It is notoriously difficult to establish the presence of a periodic signal in data dominated by red noise.  This lesson has recently been reinforced in ongoing searches for binary supermassive black holes, where promising evidence for periodicity has often weakened with additional data (e.g., \citealt{2018ApJ...859L..12L,2023MNRAS.518.4172D}; see \citealt{2016MNRAS.461.3145V} for a general discussion of false periodicities).  One of the reasons for the difficulty, which applies equally well to GRB data, is that the red noise itself can have structure which can be mistaken for periodicity.  In this section we discuss our approach, which allows the red noise to have a wide variety of shapes, and show that even with this flexibility the 19.5~Hz QPO stands out.

Our analysis follows the method of QPO detection described first in \citet{2019ApJ...871...95M}, in the context of a search for QPOs in the tail of the giant flare from the soft gamma-ray repeater SGR~1806$-$20, where there is also significant red noise.   The method was then used in \citet{2023Natur.613..253C} to discover kilohertz QPOs in Burst and Transient Source Experiment (BATSE) data on GRB~910711 and GRB~931101B.  In brief, the method performs Bayesian model comparison between a model without a QPO (which could have excess red, white, or blue noise) and one with one or more Lorentzian QPOs (which can also have excess noise), using power spectral data.  Here, within our QPO model, we also encompass the possibility of a periodic oscillation, which in practice means a QPO with an unresolvably small frequency width.  As indicated in Section~\ref{sec:lightpower}, we use the \citet{1975ApJS...29..285G} power spectrum normalization, in which the mean power is 1 from an intrinsically constant signal with only Poisson noise.  

As is evident from Figure~\ref{fig:powerspectra}, our segments, and indeed most segments of most GRBs, have substantial red noise at the low frequencies of interest in our analysis.  We emphasize that this is real, physical, variation; using the \citet{1975ApJS...29..285G} normalization the chance probability of a power $P>P_0$ from purely Poisson noise with no intrinsic variability is $e^{-P_0}$.  Thus in practice powers larger than a few tens are not produced by Poisson fluctuations.  Note that in the formalism of \citet{1975ApJS...29..285G}, if there is nonzero signal then the probability distribution of observed power is a series expansion (equation (15) from \citealt{1975ApJS...29..285G}) rather than a simple exponential.

However, we are focused not on the general continuum of red noise but on the possibly special implications of a QPO, which could point to a characteristic frequency in the system.  With that in mind, the strong excess at $\approx 19.5$~Hz in both Swift BAT and Fermi GBM data, flanked by much lower powers on either side, is worthy of investigation.

\begin{figure}
  \begin{minipage}[c]{0.65\textwidth}
    \includegraphics[width=\textwidth]{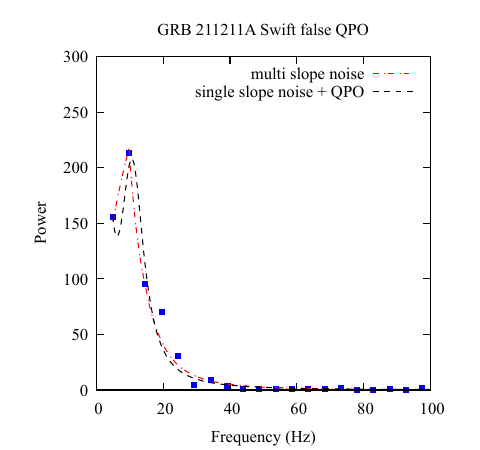}
  \end{minipage}\hfill
  \begin{minipage}[c]{0.32\textwidth}
    \caption{
Example of a power spectrum from a segment of Swift BAT data (solid blue squares) from GRB~211211A which registers as an extremely strong QPO when compared with a single-slope red noise model.  The large magnitudes of the powers at low frequencies, combined with the local maximum in power at $\approx 9.7$~Hz, gives a Bayes factor $>10^{31}$ in favor of a QPO (dashed black line) versus the single-slope red noise model.  For this reason we chose to employ a more flexible red noise model with multiple slopes at low frequencies (dashed red line), which provides an adequate fit without requiring a QPO.} \label{fig:falseQPO}
  \end{minipage}
\end{figure}

To pursue our analysis we need to decide on a red-noise-only model to compare with a model that has a QPO.  In our initial analysis we used red noise described by a single power law: $P(f)\propto f^{-\alpha}$, where $\alpha$ could range between $\alpha=-1$ (which is thus actually blue noise) and a fairly steep red noise slope of $\alpha=+3$.  But in the long and complex light curve of GRB~211211A there are segments with power spectra such as that featured in Figure~\ref{fig:falseQPO}.  In this segment, the power at the second-lowest frequency is higher than the power at the lowest frequency, and the powers are large enough that a QPO model is favored overwhelmingly compared with a single-slope red noise only model (Bayes factor $>10^{31}$). 

Although the increase in power to the second-lowest frequency is formally significant, we elect to employ a more flexible red noise model to ensure that local maxima in the power need to stand out substantially from background red noise.  Note that for our segment length of 0.2048 seconds, which was inspired by the report from \citet{2022arXiv220502186X} of a $\approx 22.5$~Hz QPO in a $\sim 0.2$~second portion of the precursor, the frequency resolution is $1/0.2048~{\rm s}\approx 4.88$~Hz and thus the initial several frequencies are $4.88$~Hz, $9.77$~Hz, etc.  The priors on our models are displayed in Table~\ref{tab:priors}.

\begin{deluxetable}{cc}
\caption{Priors on Power Spectral Models}
\tablehead{\colhead{Quantity} & \colhead{Prior (flat in indicated range)}
}
\startdata
$A_{\rm noise}(4.88~{\rm Hz})$ & 0 to 2000\\
Slope ($4.88-9.77$~Hz)  & $-1$ to 3\\
Slope ($9.77-14.65$~Hz) & $-1$ to 3\\
Slope ($14.65-19.53$~Hz)& $-1$ to 3\\ 
Slope ($19.53-24.41$~Hz)& $-1$ to 3\\ 
$A_{\rm QPO}$& 0 to 200\\ 
$\log_{10} \nu_{\rm QPO} ({\rm Hz})$& $1.0$ to $3.7$\\ 
$\log_{10} \Delta\nu_{\rm QPO} ({\rm Hz})$& $-1$ to 3
\enddata
\tablecomments{Priors on our noise and noise+QPO models for the power spectra that we analyze.  All quantities have flat priors in the indicated range, and the quantities with QPO subscripts are only used in the noise+QPO model.}
\label{tab:priors}
\end{deluxetable}

We then compute the Bayes factor ${\cal B}$ between the two models using the standard Bayesian prescription:

\begin{equation}
    {\cal B}_{\rm QPO,red}=\frac{\int {\cal L}(d|{\vec\theta}_{\rm QPO})q({\vec\theta}_{\rm QPO})d{\vec\theta}_{\rm QPO}}{\int {\cal L}(d|{\vec\theta}_{\rm red})q({\vec\theta}_{\rm red})d{\vec\theta}_{\rm red}}\; .
\end{equation}
Here ${\vec\theta}_{\rm QPO}$ represents the vector of parameters for the QPO model, ${\vec\theta}_{\rm red}$ represents the vector of parameters for the red noise only model, $q$ is the (normalized) prior, and ${\cal L}$ is the likelihood of the data $d$ given the model.  We assume that prior to our analyzing the data the models are equally probable, which means that the odds ratio ${\cal O}_{\rm QPO,red}$ between the models equals ${\cal B}_{\rm QPO,red}$.

\begin{figure}
    \centering
    \includegraphics[width=\linewidth]{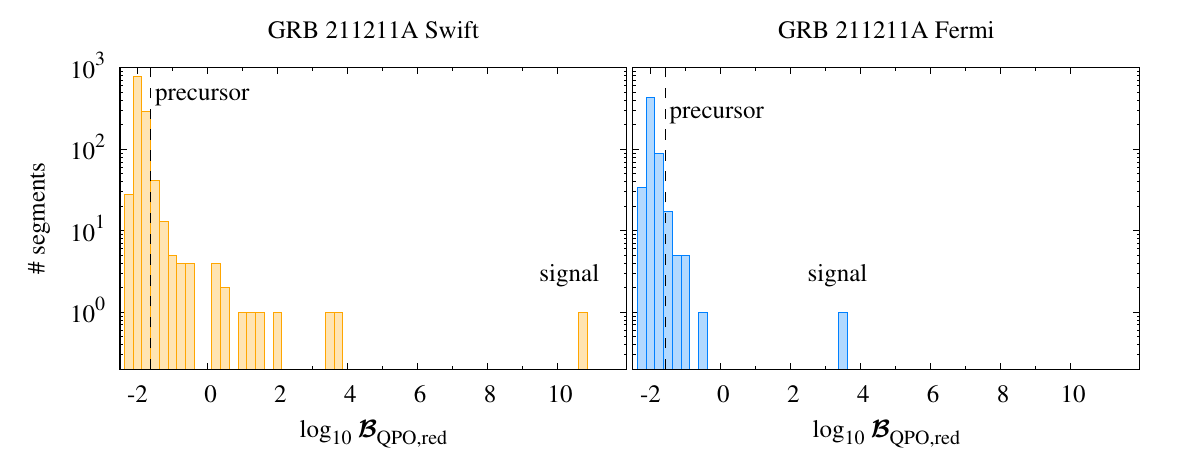}
    \caption{Differential distribution of $\log_{10}$ Bayes factors between models with red noise plus a QPO and models with just red noise (see text for details) for half-overlapping 0.2048-second segments of Swift BAT data (left panel) and Fermi GBM data (right panel) from GRB~211211A.  In each panel our featured segment in the main burst with a 19.5 Hz QPO is highlighted, and the vertical dashed line indicates the Bayes factor for the precursor segment used by \citet{2022arXiv220502186X} to suggest the presence of a $\approx 22.5$~Hz QPO.  The evidence for a signal in our featured segment stands out overwhelmingly and independently in the Swift BAT and the Fermi GBM data.  In contrast, although the Bayes factor for the precursor segment is above average in both data sets, the evidence for a QPO is not especially strong.}
    \label{fig:Bayesdist}
\end{figure}

We divide the Swift BAT data, and independently the Fermi GBM data, into segments of duration 0.2048~seconds ($=2^{11}$ times our time resolution of 0.0001~seconds), with consecutive segments overlapping by half their duration, i.e., by 0.1024~seconds (to reduce the probability that a short-lived signal will be missed).  This results in 1171 segments of Swift BAT data and 579 segments of Fermi GBM data.

Figure~\ref{fig:Bayesdist} shows the resulting distribution of Bayes factors in Swift BAT and Fermi GBM data, and highlights those for our featured segment (${\cal B}=6.9\times 10^{10}$ for Swift BAT and ${\cal B}=4.5\times 10^4$ for Fermi GBM) and the precursor (vertical dashed line).

\begin{deluxetable}{cccc}
\caption{Summary of Best Fits and Bayes Factors}
\tablehead{\colhead{Detector} & \colhead{$\nu$(Hz)} & \colhead{$\Delta\nu$(Hz)} & Bayes Factor
}
\startdata
Swift BAT & 19.5 & 0.15 & $6.9\times 10^{10}$ \\
Fermi GBM & 19.4 & 0.12 & $4.5\times 10^4$
\enddata
\tablecomments{Best fits and Bayes factors for our featured segment in the Swift BAT and Fermi GBM data.  The centroid frequency $\nu$ and frequency width $\Delta\nu$ of the fitted Lorentzian QPOs are consistent between the two data sets, and the large Bayes factors compared with a noise-only model indicate that the signal is strong for both data sets independently.}
\label{tab:bestfits}
\end{deluxetable}

The evidence for a signal in our segment stands out overwhelmingly, in both the Swift BAT and the Fermi GBM data, compared with any other segment.  The evidence is stronger from the Swift BAT data than from the Fermi GBM data, due to the larger number of counts, but in both data sets independently the signal is strong (see Table~\ref{tab:bestfits} for a summary of the best fits and Bayes factors).  

We also checked that our model provides an acceptable description of the data.  Using the formulae of \citet{1975ApJS...29..285G} we generated numerous synthetic data sets from our best-fit noise+QPO models and computed the log likelihood of the synthetic data (up to 100 Hz, i.e., the first 20 frequencies).  The log likelihood of the Swift BAT data is at the 8th percentile, and of the Fermi GBM data is at the 30th percentile, of the corresponding sets of synthetic log likelihoods.  Thus our model has captured the essential features of the low-frequency portions of the power spectra. 

It is therefore clear that for these data sets the red noise plus QPO model that we employ fits the data far better than the red noise only model.  It is, however, difficult to judge whether this is the correct red noise model, and whether it would be reasonably common for the natural high-amplitude variability of GRBs to counterfeit a signal similar to what we see in GRB~211211A.

To provide an independent measure of the significance of the signal we would like to use a model without QPOs, generate synthetic light curves with that model, and then compare the results with the data.  We lack a physical picture with which to select such models.  We therefore follow the guidance of \citet{2022ApJ...936...17H} and use Gaussian processes.  More specifically, we use as a smooth light curve model a triangular shape with a slow rise and a faster decline (which is roughly similar to the average light curve in the right hand panels of Figure~\ref{fig:lightcurves}).  A least-squares fit of that functional form gives the following for counts per 0.0064-second interval (where $t$ is in units of seconds and $t=0$ is the start of the segment):

\noindent Swift BAT:
\begin{equation}
\begin{array}{rl}
    t&\leq 0.1674~{\rm seconds}: {\rm counts}=547.4+2426.674t\\
    t&>0.1674~{\rm seconds}:  {\rm counts}=953.625-14794.124(t-0.1674)\; .\\
\end{array}
\end{equation}
Fermi GBM:
\begin{equation}
\begin{array}{rl}
    t&\leq 0.1668~{\rm seconds}: {\rm counts}=284.622+943.776t\\
    t&>0.1668~{\rm seconds}:  {\rm counts}=442.044-6409.395(t-0.1668)\; .\\
\end{array}
\end{equation}
We then sampled from a Gaussian process and added the result to this overall shape (see \citealt{2022ApJ...936...17H} for a discussion of Gaussian processes in this context).  We used a squared exponential kernel such that the covariance for two samples separated by time $\tau$ is
\begin{equation}
    \kappa(\tau)=\sigma^2\exp{[-\tau^2/2\ell^2]}\; ,
\end{equation}
with parameters $\sigma$ (the overall scale of the variance) and $\ell$ (the duration over which correlations decline).  Based on an approximate fit to the covariances of the Swift BAT data, we chose $\sigma=80$ and $\ell=0.05$~seconds, and from a fit to the Fermi GBM data we chose $\sigma=40$ and $\ell=0.1$~seconds. 

\begin{figure}
    \centering
    \includegraphics[width=\linewidth]{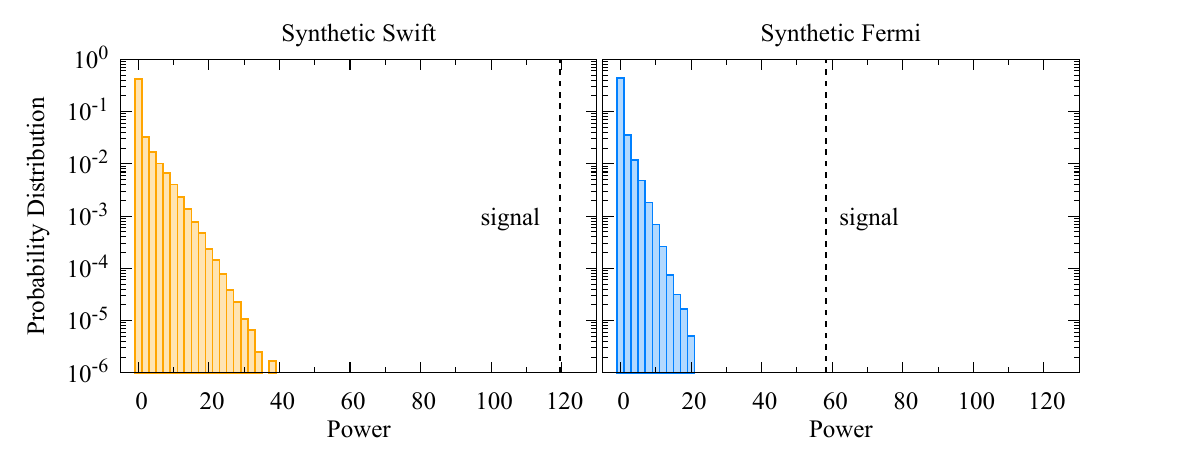}
    \caption{Differential distribution of powers at $f\geq 19.5$~Hz for light curves generated using Gaussian processes (see text for details) for synthetic light curves similar to the Swift BAT data (left panel) and the Fermi GBM data (right panel), for 50,000 realizations each.  The vertical dashed line in each panel shows the observed power in the 19.5 Hz QPO. For both the Swift BAT and the Fermi GBM data sets on our segment, the power is much larger than what emerges in our synthetic data sets.}
    \label{fig:synthpowerdist}
\end{figure}

We then generated 50,000 synthetic light curves for Swift BAT, and 50,000 synthetic light curves for Fermi GBM, using these Gaussian processes.  As a check that our approach gives red noise levels comparable to what we see in the data, we note that the observed Swift BAT powers at 4.9~Hz and 9.8~Hz are, respectively, at the 44th and 53rd percentiles of the powers at those frequencies in the synthetic Swift BAT data, and the observed Fermi GBM powers at those frequencies are, respectively, at the 45th and 95th percentiles at those frequencies in the synthetic Fermi GBM data.  The power distributions are thus roughly consistent with what we see in the observations.

Figure~\ref{fig:synthpowerdist} shows the cumulative distribution of the synthetic powers seen at a frequency of 19.5~Hz or higher, and compares that distribution with what is observed.  For both the Swift BAT and the Fermi GBM data sets, the observed power is much larger than any power seen in the synthetic data.  The distribution of powers in the synthetic data sets is not clear, but at the highest powers an exponential distribution appears roughly consistent with the data.  Using this extrapolation suggests that perhaps a billion times as many samples would be needed for there to be a good chance of obtaining powers as large as are observed in either data set.

Because we do not have a well-understood physical model for the details of the light curve of GRB~211211A or other GRBs, our quantitative results cannot be considered definitive.  However, it does suggest that signals as strong as we see are not easily produced even given large variations in the count rate.

Our final clue regarding this signal is that it starts and ends abruptly.  We see this in the dynamical power spectrum Figure~\ref{fig:spectrogram}, where the 19.5~Hz power is large only for a brief time.  We can also see this from the Bayes factors: ${\cal B}_{\rm QPO,red}\approx 6.9\times 10^{10}$ for our featured segment in the Swift BAT data, but for the previous segment (which we recall overlaps half of our featured segment) ${\cal B}_{\rm QPO,red}=0.32$, and for the following segment (also half-overlapping), ${\cal B}_{\rm QPO,red}=0.043$.  

We draw the following conclusions from the results of this section:

\begin{enumerate}
    \item There is strong red noise in our featured segment of GRB~211211A, and in many other segments from this burst.

    \item However, the power at $\approx 19.5$~Hz, lasting for 0.2048 seconds and starting roughly 2.66 seconds after the burst trigger, plus the much lower powers at the next lowest and at the next highest frequencies in our power spectra, makes this segment stand out from any other in the burst. The exact significance of the feature depends on the model of red noise, but the Bayes factor of a QPO model relative to a red noise model is orders of magnitude greater than it is for any other segment.

    \item The 19.5~Hz signal is seen strongly, and independently, in the Swift BAT and in the Fermi GBM data, and has similar characteristics (e.g., frequency and frequency width).  Thus the signal is very unlikely to be an instrumental artifact.

    \item The 19.5~Hz signal is narrow: there is negligible excess power at $\pm 5$~Hz compared with the main signal.  

    \item The 19.5~Hz signal has a higher fractional amplitude at higher photon energies in both data sets.

    \item The 19.5~Hz feature starts and ends abruptly; the full duration of the signal is not much longer than the $\approx 0.2$~seconds of our segment.

    \item The 22.5~Hz QPO suggested by \citet{2022arXiv220502186X} to exist in the burst precursor does not, in our analysis, appear to be especially significant.
\end{enumerate}

It is clear, based on the magnitudes of the powers, that the apparent 19.5~Hz signal has an astrophysical origin rather than being caused by instrumental effects or statistical fluctuations.  It is less certain that the feature we discovered indicates the presence of a narrow, coherent frequency.  However, in the next section we will proceed under the assumption that during the short duration of the signal, 19.5~Hz is characteristic, and ask what its cause might be.

\section{Implications and explanations for the QPO}
\label{sec:discussion}

From the previous section, we found that (1)~there is a strong $\approx 19.5$~Hz signal in (2)~a short ($\approx 0.2$~seconds) interval of GRB~211211A, which (3)~is much narrower than the $\approx 5$~Hz resolution of our power spectra and (4)~has higher fractional amplitude at higher photon energies.  Assuming that this is a characteristic frequency of the system which is evident for only a short time, what are some possible physical causes?

Of the QPO features listed above, the one that is likely to be the easiest to explain in the widest variety of models is the increase of fractional amplitude with increasing photon energy.  Any model with a periodically changing spectrum that has a steeply decreasing flux at higher energies will show this behavior.  For example, if the temperature $T$ of a blackbody changes periodically then the fractional amplitude of the modulation at energies many times $kT$ will be much larger than the fractional amplitude at energies $\sim kT$.  Thus although this observed feature of our QPO might be considered a rough confirmation of the physical reality of the feature, it does not discriminate between models.

We thus instead begin by considering what sources can produce frequencies of order 19.5~Hz.  The characteristic frequency of an object of average density ${\bar\rho}$ is $\sim(G\bar\rho)^{1/2}$.  Because 19.5~Hz is well above the $<1$~Hz maximum for white dwarfs and less dense objects, these are ruled out (see the similar discussion in \citealt{1968Natur.218..731G} for why pulsars cannot be white dwarfs).  Thus, even independently of GRB~211211A being a gamma-ray burst, the 19.5~Hz QPO points to a neutron star or black hole origin.

The ringdown frequency of a black hole is $\sim 10^4~{\rm Hz}(M_\odot/M)$, multiplied by a factor of order unity that depends on the black hole spin parameter and the harmonic/overtone of the mode.  Thus a $\sim 500~M_\odot$ black hole would have a frequency in the vicinity of our signal.  The observed quality factor of $Q = \pi f/\Delta f \approx \pi 20~{\rm Hz}/2~{\rm Hz} \approx 30$ is relatively high for a black hole ringdown, but would be possible if the spin parameter is $\gtorder 0.98$ (e.g., \citealt{1989PhRvD..40.3194E}).  However, the observation of a kilonova from this GRB, which suggests that a neutron star was disrupted, is not consistent with such a high-mass black hole because a neutron star would enter the horizon without being torn apart.  It therefore appears that the signal originated from a mode or rotation of a neutron star, or from some aspect of accretion disk around either a neutron star or a black hole.

Neutron star p-modes, including the fundamental f-mode, are much too high-frequency ($>1000$~Hz) to explain our signal.  Neutron star g-modes are lower in frequency but are expected to be at least hundreds of Hz and are thus also too high in frequency.  The frequencies could be lower for a proto neutron star because of its lower density, but this state is expected to evolve rapidly in density and thus it is difficult to understand how it would produce as sharp a frequency as our signal.

QPOs with frequencies comparable to the 19.5~Hz signal have been seen in giant flares from the soft gamma-ray repeaters SGR~1900+14 and SGR~1806$-$20 \citep{2005ApJ...628L..53I,2005ApJ...632L.111S,2006ApJ...653..593S,2014ApJ...793..129H,2018A&A...610A..61P,2019ApJ...871...95M}, and the frequency width is often comparable to the $\sim 2$~Hz we infer for the 19.5~Hz signal \citep{2019ApJ...871...95M}.  There is not a clear consensus about the origin of these SGR QPOs, but candidates include torsional modes of the crust and magnetohydrodynamic (MHD) modes in the core.  A challenge to crustal models of our signal is that because the oscillation is evident less than two seconds after the start of the main burst, it is implausible that a hard crust would have formed.  MHD modes are not as easy to disprove, although for this and for other frameworks there remains the question of why the signal starts and stops abruptly.  

A neutron star could rotate at a frequency compatible with our signal.  The initial rotation rate after merger would be high, in the vicinity of $\sim 1500$~Hz, which means that it would need to slow down within $\sim 1.6$ seconds to 19.5~Hz.  Candidate mechanisms for the slowdown include pulsar-like vacuum magnetic dipole radiation, interaction of a stellar magnetic field with matter falling back onto the remnant, and gravitational radiation from an asymmetric star (see, e.g., \citealt{1983bhwd.book.....S} for the relevant formulae).  We find that even for a star of ellipticity unity, gravitational radiation would take tens of thousands of seconds to spin a star down to 19.5~Hz, so this is insufficient.  The mechanisms involving magnetic field both need field strengths $\sim 10^{18}$~G to work in 1.6 seconds, which is two orders of magnitude larger than has been inferred from any other star, but might not be impossible.

However, the strongest argument against this scenario is that the energy released due to spindown is larger by a factor of several than even the isotropic equivalent energy release for GRB211211A.  \citet{2021GCN.31230....1M} estimate a total isotropic equivalent energy release from 1~keV to 10~MeV of $1.16\times 10^{52}$~erg.  High-density equations of state which sustain masses $>2~M_\odot$ have maximum-mass moments of inertia $I\sim 2\times 10^{45}$~g~cm$^2$ (e.g., \citealt{1994ApJ...424..823C}).  At an angular frequency $\Omega=2\pi\times 1500$~rad~s$^{-1}$, the rotational energy is $E_{\rm rot}=\frac{1}{2}I\Omega^2\approx 9\times 10^{52}$~erg.  Thus if the star spun down to 19.5~Hz, the fluence we would see would be much larger than what was observed from GRB~211211A.

Another possibility is free precession of the merger remnant, if it is not rotating around one of its principal axes (we thank Zorawar Wadiasingh for suggesting this possibility).  For an oblate star, the precession frequency is roughly the rotation frequency multiplied by the fractional difference in the moments of inertia (see \citealt{1970ApJ...160L..11G,1970Natur.225..838R} for early discussion of neutron star precession).  This would imply an oblateness on the order of $\sim 1-2$\%, which seems plausible.  Physically, if the direction of the jet is modified by precession then the observed flux could be modulated at this frequency.

The last possibility involves an accretion disk.  It has been suggested (e.g., \citealt{2013PhRvD..87h4053S,2023arXiv230806151L}) that if a rapidly-rotating black hole tidally disrupts a neutron star, and if the resulting accretion disk has an axis that is not aligned with the black hole rotation axis, then at high accretion rates Lense-Thirring precession could drive the disk to solid-body precession \citep{2005ApJ...623..347F} which would have a frequency in the $\sim 10-100$~Hz range.  The modulation we see could be due to precession of a jet aligned with the disk axis.  If this is the explanation, then it suggests that the black hole had low mass (because otherwise the neutron star would not be disrupted outside the horizon) and high enough spin to produce strong Lense-Thirring precession.

One of the most significant challenges to any model of the 19.5~Hz signal is to explain how it starts and then ends abruptly.  We could speculate that, for example in the precessing disk idea, it takes a certain amount of time for the disk to lock into solid-body rotation; prior to that, there would not be a clear direction to the disk axis and thus no definite frequency.  Once the disk is in solid-body rotation, it could be that alignment with the black hole rotation axis and/or rapid draining of the disk into the black hole drops the amplitude quickly.  Another consideration could be optical depth: perhaps the system needed to clear out some amount of matter before the QPO could be observed.  A full explanation almost certainly will require convincing numerical simulations, which are beyond the scope of this paper.  

\section{Conclusions}
\label{sec:conclusions}

We have presented evidence for a strong 19.5~Hz signal in the Swift BAT data, and independently in the Fermi GBM data, for the gamma-ray burst GRB~211211A.  Although this burst lasted for more than a minute, other characteristics (most notably the evidence for an associated kilonova: \citealt{2022Natur.612..223R,2022Natur.612..228T,2022Natur.612..232Y}) suggests that it was a prolonged burst after the merger of two compact objects, rather than resulting from the core collapse of a massive star.

The oscillation is evident only in a $\sim 0.2$~second interval beginning $\approx 1.6$ seconds after the start of the main burst.  The signal is also very narrow in frequency, with a width that is significantly less than 5~Hz, and its fractional amplitude increases with increasing energy in both the Swift BAT and the Fermi GBM data sets.

Of the models we considered, precession seems most consistent with the observed features.  One possibility is Lense-Thirring precession of a remnant accretion disk after the disruption of a neutron star by a low-mass and rapidly-spinning black hole (e.g., \citealt{2013PhRvD..87h4053S,2023arXiv230806151L}).  This would involve a black hole with a mass and spin that might not be represented in the current gravitational wave samples \citep{2019PhRvX...9c1040A,2021PhRvX..11b1053A,2021arXiv211103606T}.  Another possibility, which does not seem to have been explored in this context, is free-body precession of the merger remnant, which in that case would not have collapsed to a black hole by the time the QPO is evident.  In either case, targeted numerical simulations will be needed to determine whether a compact object coalescence could produce the behavior that we see, and in particular to produce a coherent signal which lasts for only a short time.

\acknowledgements

We thank Chuck Horowitz, Konstantinos Kalapotharakos, Stephen Lasage, David Radice, Jeff Scargle, and Zorawar Wadiasingh for discussions.  C. C. acknowledges support by NASA under award number 80GSFC17M0002. M. C. M. was supported in part by NASA ADAP grants 80NSSC20K0288 and 80NSSC21K0649. S.D. was supported by NASA under award number 80NSSC22K1516. This work was partially conducted at the Aspen Center for Physics, which is supported by National Science Foundation grant PHY-1607611.

\bibliography{bibfile}

\end{document}